\documentstyle[12pt]{article}
\begin{document}
\date{}
\title{Multi-particle States from the Effective Action for
Local Composite Operators: Anharmonic Oscillator}
\author{Anna Okopi\'nska\\
 Institute of Physics, Warsaw University, Bia\l ystok Branch,\\
 Lipowa 41, 15-424 Bia\l ystok, Poland}
\maketitle
\thispagestyle{empty}
\begin{abstract}
\noindent The effective action for the local composite operator
$\Phi^2(x)$ in the scalar quantum field theory with
$\lambda\Phi^4$ interaction is obtained in the expansion in
two-particle-point-irreducible (2PPI) diagrams up to five-loops. The
effective potential and 2-point Green's functions for elementary
and composite fields are derived. The ground state energy as
well as one- and two-particle excitations are calculated for
space-time dimension $n=1$, when the theory is equivalent to the
quantum mechanics of an anharmonic oscillator. The agreement
with the exact spectrum of the oscillator is much better than
that obtained within the perturbation theory.
\end{abstract}
\newpage
\section{Introduction}
The formalism of the effective action (EA)~\cite{Jac} is usually
used in relativistic quantum field theory; hovewer the
formulation is universal and provides an effective approach to
any quantum theory. Here we consider the theory of a real scalar
field in $n$-dimensional Euclidean space-time with a classical
action given by
\begin{equation}
S[\Phi]=\int\![\frac{1}{2}\Phi(x)(-\partial^2+m^2)\Phi(x)+
\lambda\Phi^{4}(x)]\,d^{n}x.
\label{Scl}
\end{equation}
The simplest case of $n=1$ dimensional space-time, which is
equivalent to the quantum mechanics of the anharmonic oscillator
(AO), is frequently used as a testing ground for various
field-theoretical methods. Here we shall discuss the method of
the EA for local composite operators which provides a systematic
approximation scheme for vacuum energy and lowest multi-particle
excitations. We shall keep the dimension of the space-time $n$
arbitrary as long as possible, seting $n=1$ only in the last
stage, where the energies are calculated.

The conventional EA, which is a generating functional for
one-particle-irreducible (1-PI) Green's functions, is obtained
by introducing a source coupled to the quantum field $\Phi(x)$.
By coupling external sources to bilocal $\Phi(x)\Phi(y)$~\cite{CJT} and
local $\Phi^2(x)$~\cite{Fuk} fields, the generating functionals for
composite operators are defined~(for review see Ref.\cite{Hay}).
For an interacting theory the exact form of any functional is
not known, so one resorts to approximations. Formulating an
approximation scheme for a generating functional, a consistent
set of approximate Green's functions can be obtained through
differentiation. This is crucial for a relativistic quantum
field theory, where the process of renormalization has to be
performed. Any functional, if calculated exactly, contains the
same information and gives the same result for a physical
quantity; however, the same approximation scheme for various
functionals would result in different approximations of Green's
functions and observables. Therefore, an appropriate choice of a
generating functional for calculating a quantity of interest is
important. The conventional EA is used for discussing a vacuum
structure and one-particle excitations. For a simultaneous study
of two-particle excitations, the EA for composite operators is
more suitable, since it determines the conventional EA (by
eliminating the expectation values of composite operators) and
generates Green's functions related directly to one- and
two-particle eigenmodes.

Generating functionals can be calculated in the loop expansion.
The conventional EA, $\Gamma[\varphi]$, is given by a sum of
one-particle irreducibile vacuum diagrams~\cite{Jac}. The EA for
the bilocal composite operator, $\Gamma[\varphi(x),G(x,y)]$ is
given by two-particle-irreducibile (2PI) diagrams~\cite{CJT}.
The one-loop result, after eliminating the full propagator
$G(x,y)$, gives the Gaussian approximation for the conventional
EA; however, beyond one-loop the gap equation for $G(x,y)$ is a
highly non-trivial integral equation. The EA for the local
composite operator, $\Gamma[\varphi(x),\Delta(x)]$, can be also
obtained diagramaticaly~\cite{chin,ver} and the one-loop result
gives the Gaussian approximation. Calculations of post-Gaussian
corrections are easier in this approach, since the gap equation
for the vacuum expectation value of the local composite field is
algebraic.

The vacuum functional for the composite operator $\Phi^2(x)$ is
represented by a path integral
\begin{equation}
Z[J_{1},J_{2}]=e^{W[J_{1},J_{2}]}=\int\! D\Phi\,e^{-S[\Phi] +
\int\!J_{1}(x)\Phi(x)\,d^{n}x
+\frac{1}{2}\int\!J_{2}(x)\Phi^{2}(x)\,d^{n}x}
\label{Z}
\end{equation}
and the EA is obtained as a Legendre transform
\begin{equation}
\Gamma[\varphi,\Delta]=W[J_{1},J_{2}]-\int\! J_{1}(x)\varphi(x)\,d^{n}x-
\frac{1}{2}\int\! J_{2}(x)(\varphi^{2}(x)+\Delta(x))\,d^{n}x,
\label{Gam}
\end{equation}
where
\begin{equation}
\frac{\delta W}{\delta J_{1}(x)}=\varphi(x),~~{\mbox and}~~
\frac{\delta W}{\delta J_{2}(x)}=\frac{1}{2}(\varphi^{2}(x)+\Delta(x))
\label{dJ}
\end{equation}
determine the expectation values of the fields $\Phi$ and $\Phi^2$,
in the presence of external currents $J_{1}$ and $J_{2}$. The EA
fulfils
\begin{equation}
\frac{\delta \Gamma}{\delta\varphi(x)}=-J_{1}(x)-J_{2}(x)\varphi(x),
\end{equation}
\noindent and
\begin{equation}
\frac{\delta \Gamma}{\delta \Delta(x)}=-\frac{1}{2}J_{2}(x).
\label{dde}
\end{equation}
Setting $J_{1}=J_{2}=0$, which reproduces the physical
theory~(\ref{Scl}), results in variational equations
\begin{equation}
\frac{\delta \Gamma}{\delta \varphi(x)}=0
\label{dfi0}
\end{equation}
and
\begin{equation}
\frac{\delta \Gamma}{\delta \Delta(x)}=0.
\label{ddel0}
\end{equation}
These equations determine the vacuum expectation values
$\varphi_{0}$ and $\Delta_{0}$, which are space-time independent,
by translational invariance.

The conventional EA can be obtained as
$\Gamma[\varphi]=\Gamma[\varphi,\Delta_{0}]$ with
$\Delta_{0}[\varphi]$ determined by inverting the gap
equation~(\ref{ddel0}). The effective potential (EP), defined by
\begin{equation}
V(\phi)=-\frac{\left.\Gamma[\varphi]\right|_{\varphi(x)=\phi
=const}}{\int\! d^{n}x},
\end{equation}
gives the vacuum energy density $V(\varphi_{0})$.

Green's functions, generated from the EA for local composite
operators, provide a convenient tool to study multi-particle
states, since their zero modes give directly the excitation
energies above the ground state. One-particle eigenmode is
determined by the 2-point Green's function for the elementary
field
\begin{equation}
\Gamma^{2}(x-y)=\left.\frac{\delta^2 \Gamma}{\delta
\varphi(x)\delta \varphi(y)}\right|_{\varphi(x)=\varphi_{0},
\Delta(x)=\Delta_{0}},
\label{g2}
\end{equation}
which is an inverse of the full propagator
\begin{equation}
W^{2}(x-y)=<T\Phi(x)\Phi(y)>_{connected}.
\label{w2}
\end{equation}
An appropriate function to study two-particle excitation is the
2-point Green's function for the composite field
\begin{equation}
\Gamma^{4}(x-y)=\left.\frac{\delta^2 \Gamma}{\delta
\Delta(x)\delta \Delta(y)}\right|_{\varphi(x)=\varphi_{0},
\Delta(x)=\Delta_{0}}
\label{g4}
\end{equation}
which is an inverse of the function
\begin{equation}
W^{4}(x-y)=<T\Phi^2(x)\Phi^2(y)>_{connected},
\label{w4}
\end{equation}
called polarisation (density fluctuaction) propagator in
many-body physics. This is an advantage of the EA for the composite
operator that $\Gamma^2$, as well as $\Gamma^4$, can be obtained
from $\Gamma[\varphi,\Delta]$ through differentiation. In the
conventional approach $\Gamma^4$ cannot be derived directly from
the EA, but the polarisation propagator $W^4$ has to be
calculated and its inverse has to be found.

In Section~\ref{2PPI} the 2PPI expansion for the effective
action for the composite operator $\Phi^2(x)$ is discussed, the
EP and Green's functions $\Gamma^{2}$ and $\Gamma^{4}$ are
obtained up to five loops for the scalar theory in the
space-time of $n$-dimensions. In Section~\ref{QM} we discuss the
method for $n=1$, when the theory has a physical interpretation
of a quantum-mechanical anharmonic oscillator. The EP and
spectral properties of Green's functions obtained in 2PPI
expansion are studied and the resulting energies are compared
with the exact spectrum of the AO. Our conclusions are
summarised in Section~\ref{con}.

\section{The 2PPI expansion}
\label{2PPI}
Verschelde and Coppens~\cite{ver} represented the EA for the
local composite operator in the form
\begin{eqnarray}
\lefteqn{\Gamma[\varphi,\Delta]=-S[\varphi]-6\lambda
\int \varphi^2(x)\Delta(x)d^{n}x-3\lambda
\int \Delta^2(x)d^{n}x}\nonumber\\
&+&\frac{1}{2} \int \Delta(x) (\Omega^2[\varphi,\Delta]-m^2)d^{n}x
+\Gamma^{2PPI}[\varphi,\Omega^2[\varphi,\Delta]],~~~
\label{G2PPI}
\end{eqnarray}
where
\begin{equation}
\Omega^2=m^2-J_{2}[\varphi,\Delta]+12\lambda (\varphi^2+\Delta).
\label{Omega}
\end{equation}
They have shown that
$\Gamma^{2PPI}[\varphi,\Omega^2[\varphi,\Delta]]$ is a sum of
the 2PPI diagrams, defined as those which stay connected after
cutting one or two internal lines meeting in the same vertex.
The (inverse) propagator is given by
\begin{equation}
G^{-1}(x,y)=(-\partial^2+\Omega^2(x))\delta(x-y),
\label{pro}
\end{equation}
where the effective mass $\Omega$ is determined as a function of
$\Delta$ and $\varphi$ by inverting the equation
\begin{equation}
\frac{\delta\Gamma^{2PPI}}{\delta \Omega^2(x)}=-\frac{\Delta(x)}{2},
\label{var}
\end{equation}
obtained from equation~(\ref{dde}). We have slightly modified the
notation of Verschelde and Coppens, by introducing the effective
mass $\Omega[\varphi,\Delta]$ and using $\lambda$ not divided by
4!

The conventional $\Gamma[\varphi]$ is given by
$\Gamma[\varphi,\Delta_{0}]$, with $\Delta_{0}[\varphi]$
determined by the gap equation
\begin{equation}
\Omega^2[\varphi,\Delta]=m^2+12\lambda(\Delta+\varphi^2),
\label{gap}
\end{equation}
obtained by setting $J_{2}=0$ in equation(\ref{Omega}). Instead
of $\Delta_{0}[\varphi]$ we shall use a self-consistent mass
$\Omega_{0}[\varphi]$, determined from Eq.~\ref{var}
and~\ref{gap}.

Green's functions in the 2PPI expansion are obtained as
functional derivatives of $\Gamma[\varphi,\Delta]$ at
$\varphi=\varphi_{0}$ and $\Delta=\Delta_{0}$. For simplicity,
we consider here the case of unbroken reflection symmetry,
taking $\varphi_{0}=0$. The 2-point Green's function for the
elementary field~(\ref{g2}) is given by
\begin{equation}
\Gamma^{2}(x-y)=-(-\partial^2+\Omega^2)\delta(x-y)+\left.\frac{\delta^2
\Gamma^{2PPI}}{\delta\varphi(x)\delta \varphi(y)}\right|_{\varphi(x)=0,
\Omega(x)=\Omega_{0}}.
\label{g2loc}
\end{equation}
and its Fourier transform will be denoted by $\Gamma^2(p)$. The
2-point Green's function for the composite field~(\ref{g4}) can be
represented by
\begin{equation}
\Gamma^{4}(x-y)=-6\lambda\delta(x-y)-\frac{1}{2}\Pi^{-1}(x-y)
\label{g4loc}
\end{equation}
where
\begin{equation}
\Pi(x-y)=2\left.\frac{\delta^2\Gamma^{2PPI}}{\delta\Omega^2(x)\delta
\Omega^2(y)}\right|_{\varphi(x)=0, \Omega(x)=\Omega_{0}}
\label{pi}
\end{equation}
is a sum over irreducible polarisation parts. Its Fourier
transform is given by
\begin{equation}
\Gamma^{4}(p)=-6\lambda-\frac{1}{2\Pi(p)}.
\label{gg4}
\end{equation}

The EA can be expanded in powers of the Planck constant $\hbar$,
which is equivalent to the expansion in the number of
loops~\cite{Jac}. As shown by Verschelde and Coppens, in the EA
for the local composite operator only the 2PPI diagrams are
present. The one-loop approximation, after setting $\hbar=1$, is
given by
\begin{eqnarray}
\lefteqn{\Gamma_{1}[\varphi,\Delta]=-\int[\frac{1}{2}\varphi(x)
(-\partial^2+m^2)\varphi(x)+\lambda\varphi^{4}(x)]\,d^{n}x}\nonumber\\
&+&\frac{1}{2}\int d^{n}x
\Delta(x)(\Omega^2[\varphi,\Delta]-m^2-12\lambda\varphi^2(x)
-6\lambda\Delta(x))-\frac{1}{2}TrLnG^{-1}~~~~~
\label{GEAloc}
\end{eqnarray}
with the effective mass equation~(\ref{var}) given by
\begin{equation}
\Delta(x)=G(x,x).
\label{var1}
\end{equation}

The conventional EA, obtained as $\Gamma[\varphi,\Delta_{0}]$
with a self-consistent mass $\Omega_{0}$, determined by the gap
equation
\begin{equation}
\Omega^{2}(x)-m^2-12\lambda\varphi^{2}(x)-12\lambda G(x,x) = 0,
\label{gap1}
\end{equation}
coincides with the Gaussian EA obtained in other approaches: in
the time-dependent Hartree approximation~\cite{JK}, from the
EA~for the bilocal composite operator~\cite{CJT}, and in the
optimized expansion~\cite{AOGF}. The Gaussian EP~\cite{GEP}
at $\phi=0$ gives the vacuum energy density in the form
\begin{equation}
V_{I}(0)=I_{1}(\Omega_{0})+\frac{1}{2}(m^2-\Omega_{0}^2)
I_{0}(\Omega_{0})+3\lambda I_{0}^2(\Omega_{0}),
\end{equation}
with $\Omega_{0}$ fulfiling an algebraic gap equation
\begin{equation}
\Omega^{2}-m^2-12\lambda I_{0}(\Omega)=0,
\label{gapIp}
\end{equation}
where
\begin{eqnarray}
I_{1}(\Omega)=\frac{1}{2}\int\!\frac{d^{n}p}{(2\pi)^{n}}\ln (p^2+\Omega ^2)
{}~~\mbox{and}~~
I_{0}(\Omega)=\int\!\frac{d^{n}p}{(2\pi)^{n}}\frac{1}{p^2+\Omega ^2}.
\end{eqnarray}
To this approximation the two-point vertex for the elementary field
\begin{eqnarray}
-\Gamma_{1}^{2}(p)=p^2+\Omega_{0}^2,
\label{proI}
\end{eqnarray}
is the same as obtained from the Gaussian EA. The 2-point vertex
for the composite field, obtained by differentiation of
$\Gamma_{1}[\varphi,\Delta]$, is equal to
\begin{equation}
-\Gamma^{4}_{1}(p)=6\lambda+\frac{1}{I_{-1}(p)}
\label{ggg4}
\end{equation}
where
\begin{equation}
I_{-1}(p)=2\int\!\frac{d^{n}q}{(2\pi)^{n}}\frac{1}
{(q^2+\Omega^2)((p+q)^2+\Omega^2)}.
\end{equation}
This is at variance with the inverse of the polarisation propagator,
which is obtained~\cite{AOGF} from the Gaussian EA in the form
\begin{eqnarray}
\lefteqn{W^{4}_{1}(p)=\frac{I_{-1}(p)}{1+6\lambda I_{-1}(p)}}\nonumber\\
&-&288\lambda^2\!\int\!\frac{d^{n}q d^{n}q'}
{[1+6\lambda I_{-1}(q+q')](q^2\!+\!\Omega^2)[(p+q)^2+\Omega^2]
(q'^2+\Omega^2)[(p+q')^2+\Omega^2]}.~~~~~~
\label{ww4}
\end{eqnarray}

Higher orders of the 2PPI expansion provide corrections to the
Gaussian approximation. The EP and vertices can be calculated
from the given order approximation of $\Gamma[\varphi,\Delta]$.
The obtained expressions can be represented in terms of Feynman
diagrams in momentum space, since $\phi$, $\Delta_{0}$ and
$\Omega_{0}$ are space-time independent. The gap equation for
$\Omega_{0}$ becomes an ordinary non-linear equation. In Fig.1
we show a diagrammatic representation for $V(0)$ and $\Pi(p)$
(up to five loops) and for $\Gamma^2(p)$ (up to four loops).
Analytical expressions can be read from the figure.

\section{Quantum-mechanical anharmonic \mbox{oscillator}}
\label{QM}
In the space-time of one dimension the $\lambda\Phi^{4}$ theory is
equivalent to the quantum mechanics of the AO with a Hamiltonian
given by
\begin{equation}
H=\frac{1}{2} p^2+\frac{1}{2} m^2 x^2+\lambda x^4.
\label{AO}
\end{equation}
The Euclidean propagators are given by
\begin{eqnarray}
W^{2}(T)&=&G_{c}(T,0)=\sum_{1}^{\infty}|\ll 0|x|k\gg |^2
e^{-|T|\varepsilon_{k}}
\end{eqnarray}
and
\begin{eqnarray}
W^{4}(T)&=&G_{c}(T,T,0,0)+2G_{c}^{2}(T,0)=
\sum_{2}^{\infty}|\ll\!0|x^2|k\!\gg |^2 e^{-|T|\varepsilon_{k}},
\end{eqnarray}
where $G_{c}$ are connected Green's functions with the number of
points equal to the number of arguments, $|k\!\gg$ denotes the
$k$th excited state of the AO~with excitation energy
$\epsilon_{k}=E_{k}-E_{0}$, and $x$ is a position operator in the
Schr\"odinger representation. The Fourier transform gives the
functions
\begin{eqnarray}
W^{2}(p)=\sum_{1}^{\infty}\frac{2\epsilon_{k}|\ll\!0|x|k\!\gg |^2}
{p^2+\varepsilon_{k}^2};~~~
W^{4}(p)=\sum_{2}^{\infty}\frac{2\epsilon_{k}|\ll\!0|x^2|k\!\gg |^2}
{p^2+\varepsilon_{k}^2}
\end{eqnarray}
which have poles at imaginary momenta, their absolute values are
equal to excitation energies. The Green's functions $\Gamma^{2}$
and $\Gamma^{4}$, generated from the exact EA for the local
composite operator, are the inverses of the above propagators.
They have an infinite number of zeros, which determine odd and
even excitations of the AO, respectively. However, in the given
order of the 2PPI expansion every Green's function has only a
finite number of zeros and gives an approximation of some part
of the energy spectrum.

The EP and Green's functions in the 2PPI expansion, read from
Fig.1, can be calculated easily in one dimensional space-time.
The powers of $\hbar$ are given explicitely in the obtained
expressions to identify orders of the expansion, in the given
order approximation we put $\hbar=1$. The value of the EP at
$\phi=0$ gives the ground state energy equal to
\begin{eqnarray}
E_{0}=-\frac{(\Omega_{0}^2-m^2)^2}{48\lambda}+\hbar\frac{\Omega_{0}}{2}
-\hbar^3\frac{3 \lambda^2}{8 \Omega_{0}^5}+\hbar^4\frac{27 \lambda^3}
{16 \Omega_{0}^8}-\hbar^5\frac{2373\lambda^4}{128 \Omega_{0}^{11}}
\label{qmep}
\end{eqnarray}
with $\Omega_{0}$ determined by the gap equation
\begin{eqnarray}
\frac{(\Omega^2-m^2)}{12\lambda}-\hbar\frac{1}{2\Omega}-
\hbar^3\frac{15\lambda^2}{8\Omega^7}+\hbar^4\frac{27\lambda^3}
{2\Omega^{10}}-\hbar^5\frac{26103\lambda^4}{128\Omega^{13}}=0.
\label{qmg}
\end{eqnarray}
The 2-point vertex for the elementary field, read from Fig.1, is
calculated to be equal to
\begin{eqnarray}
\lefteqn{-\Gamma^{2}(p)=\Omega_{0}^2 +p^2-\frac{72\hbar^2\lambda^2}
{(9\Omega_{0}^2 + p^2)\Omega_{0}^2} + \frac{108\hbar^3 \lambda^3
(45\Omega_{0}^2 +p^2)}{\Omega_{0}^5(9\Omega_{0}^2 +p^2)^2}}\nonumber\\&-&
\frac{27\hbar^4\lambda^4(589275\Omega_{0}^6 + 73413\Omega_{0}^4 p^2 +
2841\Omega_{0}^2 p^4 +47p^6)}{\Omega_{0}^8(9\Omega_{0}^2 +p^2)^3
(25\Omega_{0}^2 +p^2)}
\label{ga2}
\end{eqnarray}
and $\Gamma^{4}(p)$ can be found from~(\ref{gg4}) with the
irreducible polarisation given by
\begin{eqnarray}
\lefteqn{\Pi(p)=\frac{\hbar}{(4\Omega_{0}^2 + p^2)\Omega_{0}} +
\frac{3\hbar^3 \lambda^2 (2240\Omega_{0}^4 + 148\Omega_{0}^2 p^2 + 5 p^4)}
    {4\Omega_{0}^7(4\Omega_{0}^2 +p^2)^2 (16\Omega_{0}^2 +p^2)}}\nonumber\\&-&
\frac{27\hbar^4\lambda^3(10240\Omega_{0}^6 + 816\Omega_{0}^4 p^2 +
45\Omega_{0}^2 p^4+p^6)}
{\Omega_{0}^{10}(4\Omega_{0}^2 +p^2)^2 (16\Omega_{0}^2 + p^2)^2)}\nonumber\\&+&
\frac{3\hbar^5\lambda^4}{64\Omega_{0}^{13}(4\Omega_{0}^2 +p^2)^3
 (16\Omega_{0}^2 +p^4)^3 (36\Omega_{0}^2 + p^2)}\nonumber\\&\times&
 (266867048448\Omega_{0}^{12}+97427046400\Omega_{0}^{10}p^2 +
 11156323328\Omega_{0}^8 p^4\nonumber\\&+&783135808\Omega_{0}^6 p^6+
 35476144\Omega_{0}^4 p^8+876428\Omega_{0}^2 p^{10}+ 8701 p^{12}).
\label{pa}
\end{eqnarray}

Excitation energies are calculated as zeros of
$\Gamma-$function, determined to the given order in $\hbar$. The
root of $\Gamma^{2}(p)=0$ is calculated to be at
\begin{eqnarray}
\epsilon_{1}=\Omega_{0} -\frac{9\hbar^2\lambda^2}{2\Omega_{0}^5}+
\frac{297\hbar^3 \lambda^3}{8\Omega_{0}^8} - \frac{4599\hbar^4 \lambda^4}
{8\Omega_{0}^{11}},
\label{e1}
\end{eqnarray}
and that of $\Gamma^{4}(p)=0$ is equal to
\begin{eqnarray}
\epsilon_{2}=2\Omega_{0} + \frac{3\hbar \lambda}{\Omega_{0}^2} -
\frac{117\hbar^2 \lambda^2}{4\Omega_{0}^5} + \frac{3159\hbar^3 \lambda^3}
{8\Omega_{0}^8} -
\frac{488565\hbar^4 \lambda^4}{64\Omega_{0}^{11}}.
\label{e2}
\end{eqnarray}

Expanding the above energies to the fourth order in powers of
$\lambda$ gives
\begin{eqnarray}
\epsilon_{0}^{pert}&=&m/2 + \frac{3\lambda}{4m^2}-\frac{21\lambda^2}{8m^5} +
\frac{333\lambda^3}{16m^8} - \frac{30885\lambda^4}{128m^{11}}\nonumber\\
\epsilon_{1}^{pert}&=&m + \frac{3\lambda}{m^2} -\frac{18\lambda^2}{m^5} +
\frac{1791\lambda^3}{8m^8} - \frac{3825\lambda^4}{m^{11}}\nonumber\\
\epsilon_{2}^{pert}&=&2m + \frac{9\lambda}{m^2} - \frac{297\lambda^2}{4m^5}
+\frac{9873\lambda^3}{8m^8} - \frac{1772685\lambda^4}{64m^{11}},
\label{eep}
\end{eqnarray}
in agreement with the perturbation theory for Schr\"odinger
equation. The perturbative energies~(\ref{eep}) can be obtained
from the loop expansion of the conventional EA~\cite{AOGF};
however, more diagrams has to be evaluated and a calculation of
$\epsilon_{2}$ is furthermore complicated, since $\Gamma^{4}$
cannot be derived directly from the conventional EA. The 2PPI
expansion provides the simplest way for a field theoretical
derivation of the perturbative (in $\lambda$) results for the
ground state and two lowest excitations.

We have studied numerical results for the ground state energy
and two lowest excitations in the 2PPI expansion. To the given
order, the largest positive root of the gap equation has been
found numerically; in the case where the solution became
complex, the real part of the approximant was taken. The results
are compared with perturbative energies~(\ref{eep}) and exact
eigenvalues, calculated by the numerical procedure based on the
modification of the linear variational method~\cite{AOAO}. All
results are presented as functions of a dimensionless fraction
$z=\frac{m^2}{2\lambda^{2/3}}$, which is the only parameter of
the theory, after rescaling all quantities in terms of
$\lambda$.

The results for ground state energy~(\ref{qmep}) in successive
orders of 2PPI expansion are shown in Fig.2. The quality of
the approximation is very good in the whole range of the
parameter $z$, only for values smaller than
$z_{0}^{2PPI}\!\approx\!.2$ small discrepancies between
different orders of the 2PPI expansion and the exact result
appear. The perturbative results, differ heavily below the much
larger critical value $z_{0}^{pert}\!\approx\!2.5$. In Fig.3 we
show the first excitation energy~(\ref{e1}) in successive orders
of the 2PPI expansion, compared with the exact result. There is
the critical value $z_{1}^{2PPI}\!\approx\!1.3$, above which the
agreement is good; this value is much smaller than the critical
value for perturbative results $z_{1}^{pert}\!\approx\!3.3$. In
Fig.4 the energies~(\ref{e2}), obtained by solving
$\Gamma^{4}(p)=0$ to the given order in $\hbar$, are compared
with the second excitation energy. The agreement is worse than
in previous cases. The critical value
$z_{2}^{2PPI}\!\approx\!2.2$ is again smaller than a critical
value for perturbative results $z_{2}^{pert}\!\approx\!4.$

\section{Conclusions}
\label{con}
The 2PPI expansion for EA for the local composite operator is a
convenient tool to study a vacuum and lowest excitations in the
scalar quantum field theory. The one-loop result gives the
Gaussian approximation, where the effective propagator is a
Hartree one. Successive approximations of the EP and the Green's
functions~$\Gamma^{2}$ and $\Gamma^{4}$ have been obtained; to
each order the effective mass is determined from the algebraic
gap equation.

The results for ground state energy and two lowest excitations
have been calculated in successive orders of 2PPI expansion for
the theory in the space-time of one dimension, i.e., for quantum
mechanical AO. The gap equation has been solved numerically. The
ground state energy was obtained as a value of the EP at
$\phi=0$ and excitation energies were determined as zeros of
appropriate Green's functions. A comparison with the exact
spectrum of the AO shows that the convergence of the 2PPI
expansion is the best for the ground state. For excited states
the critical value above which the expansion converges to the
exact result is greater. The higher excitation, the larger
critical value, and the region of applicability of the
approximation diminishes. This is similar as in the perturbation
theory; however, the region of applicability of the 2PPI
expansion is much larger for all energy levels.

The numerical results of the 2PPI approach for the lowest
excitations up to second order are very similar to that obtained
in the optimized expansion and the large N expansion to the same
order~\cite{AOGF}. Even to this order, the calculations in the
later methods are not as straightforward as in the 2PPI
expansion, since $\Gamma^{4}$ cannot be derived directly. In
higher orders the optimized and large N expansion would become
even more complicated, requiring to solve a gap equation of
Bethe-Salpeter type. With the use of the
Cornwall-Jackiw-Tomboulis EA for bilocal composite operators the
situation would be very much the same. Therefore, the EA for
local composite operators provides the simplest method to study
the ground state and the lowest excitations in the case of the
AO. We hope that the application of the 2PPI expansion to the EA
for local composite operators will also appear useful for
approximate study of two-particle excitations in quantum field
theories in higher dimensional space-time.\\

\noindent {\bf{\Large Acknowledgement}}\\

\noindent This work has been supported in part by Grant
PB-2-0956-91-01 of the Committee for Scientific Research.

\newpage
\noindent {\bf{\Large Figure captions}}\\

\noindent Figure 1. The value of the EP~at $\phi=0$ and the
functions $\Gamma^2(p)$ and $\Pi(p)$ in the 2PPI expansion up to
five loops.\\

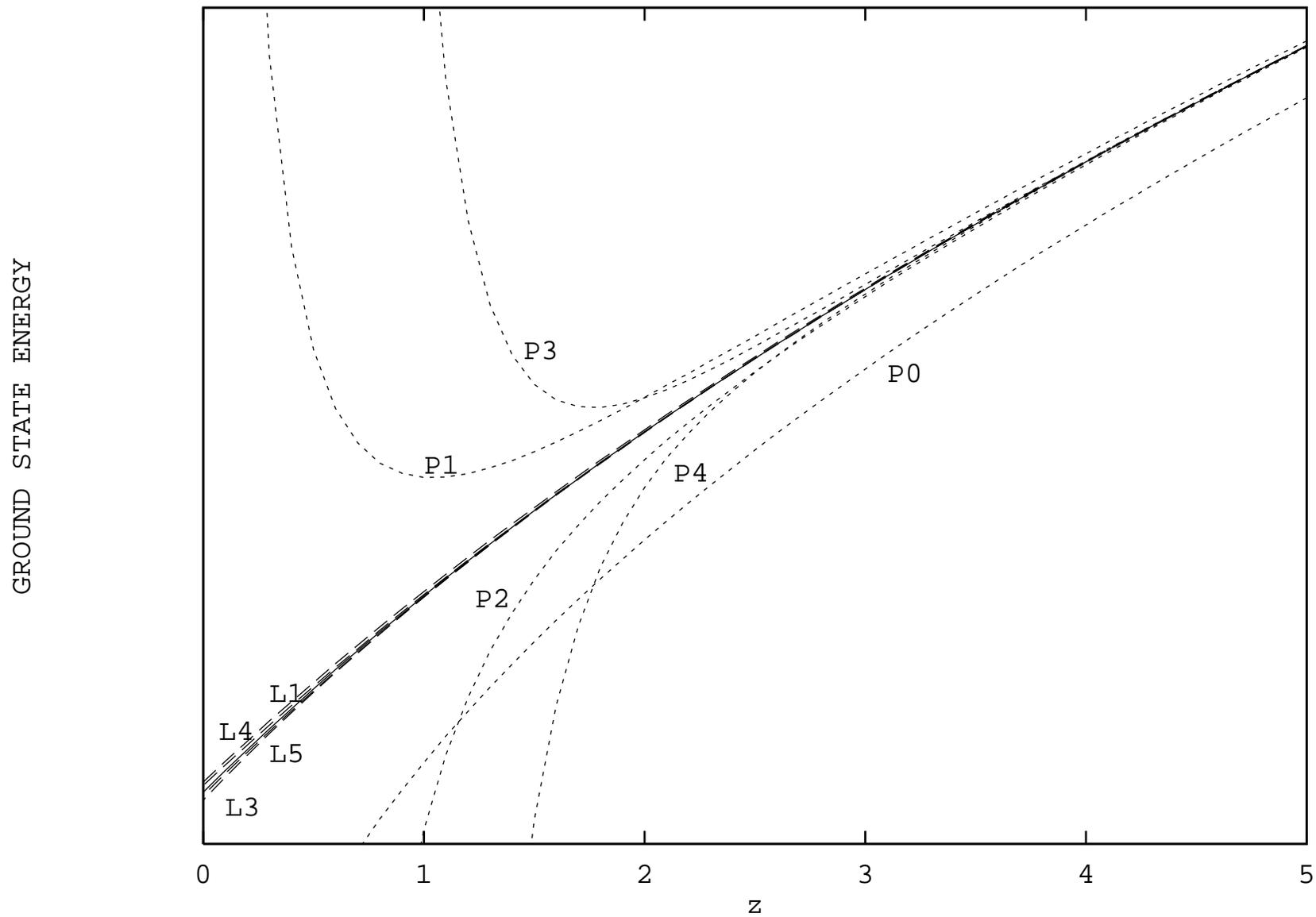
\noindent Figure 2. The ground-state energy of the AO, obtained
to the given number of loops in the 2PPI expansion {\em (dashed
lines)}, plotted {\em vs.} $z=\frac{m^2}{2\lambda^{2/3}}$;
compared with the exact value {\em (solid line)} and
given order perturbative results{\em (dotted lines)}.\\

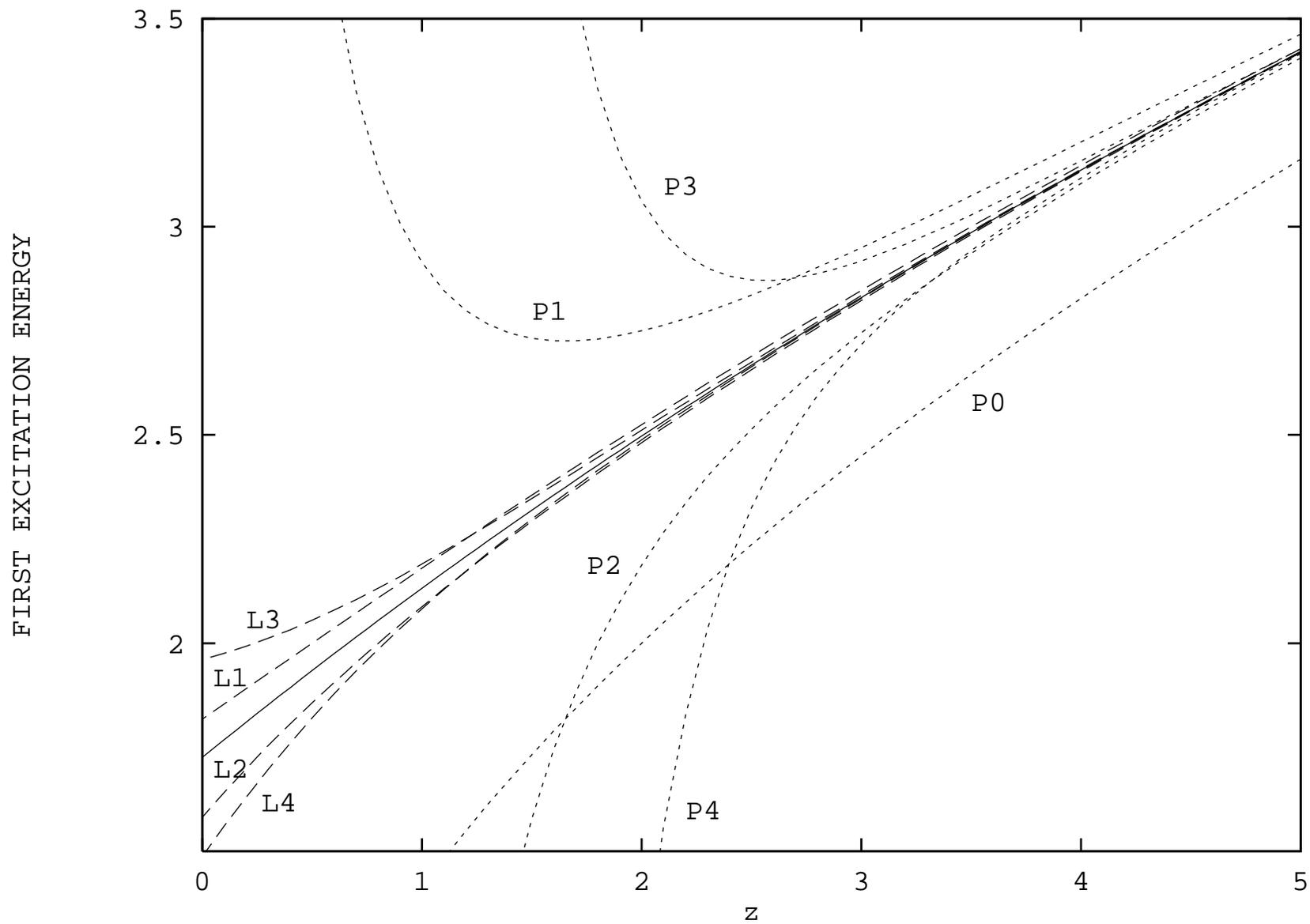
\noindent Figure 3. As in Fig.2, but for first excitation energy
of the AO, obtained as zero of $\Gamma_{2}$ function in the 2PPI
expansion.\\

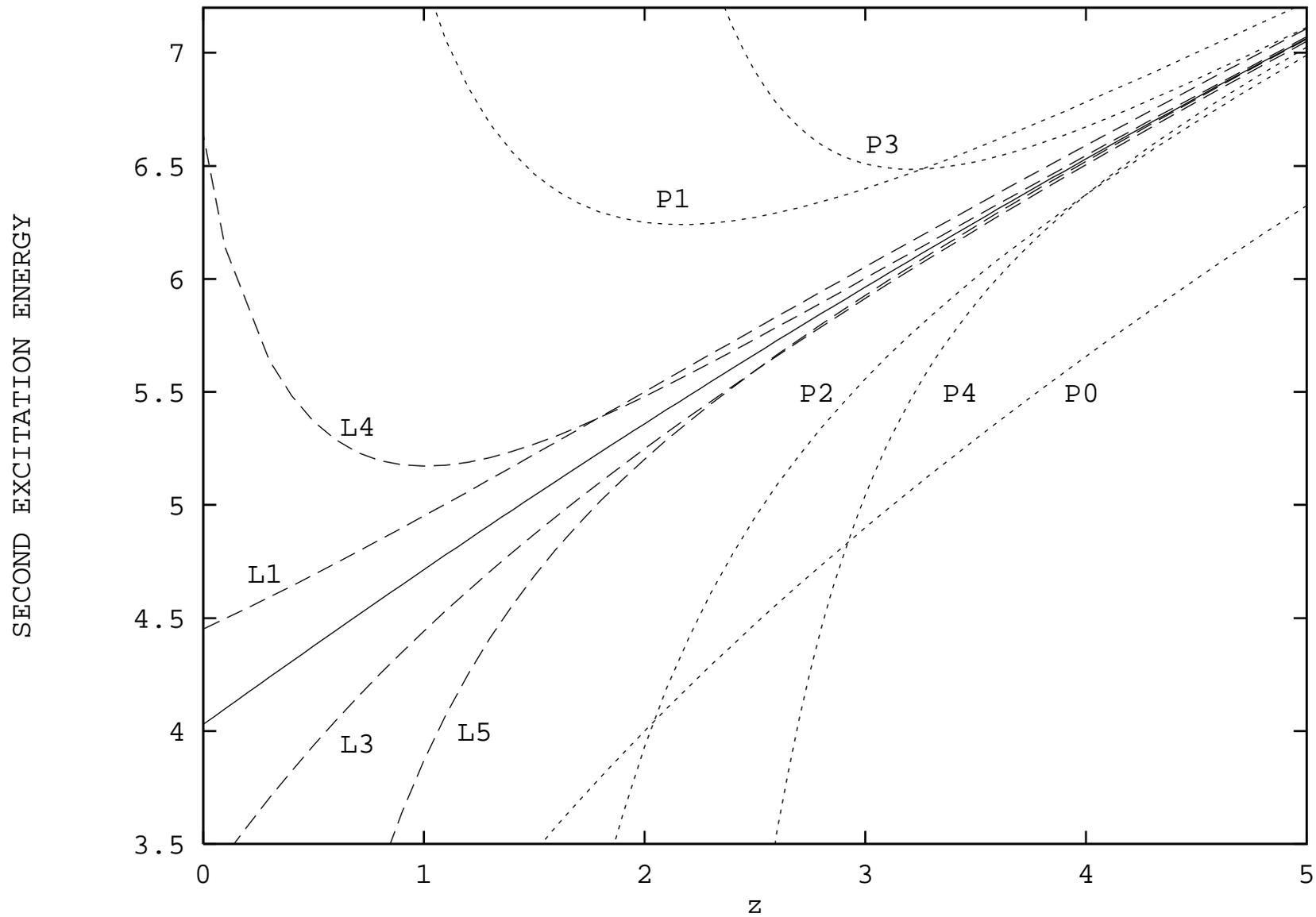
\noindent Figure 4. As in Fig.2, but for second excitation
energy of the AO, obtained as zero of $\Gamma_{4}$ function in
the 2PPI expansion.\\
\end{document}